# Radio-based Traffic Flow Detection and Vehicle Classification for Future Smart Cities

**Marcus Haferkamp[1], Manar Al-Askary[1], Dennis Dorn[2], Benjamin Sliwa[1], Lars Habel[3],
Michael Schreckenberg[3] and Christian Wietfeld[1]**

[1]Communication Networks Institute, TU Dortmund University, 44227 Dortmund, Germany
e-mail: {Marcus.Haferkamp, Manar.Askary, Benjamin.Sliwa, Christian.Wietfeld}@tu-dortmund.de
[2]Wilhelm Schröder GmbH, 58849 Herscheid, Germany
e-mail: {Dorn}@mfds.eu
[3]Physics of Transport and Traffic, University Duisburg-Essen, Germany
e-mail: {Habel, Schreckenberg}@ptt.uni-due.de

*Abstract*—Intelligent Transportation Systems (ITSs) providing vehicle-related statistical data are one of the key components for future smart cities. In this context, knowledge about the current traffic flow is used for travel time reduction and proactive jam avoidance by intelligent traffic control mechanisms. In addition, the monitoring and classification of vehicles can be used in the field of smart parking systems. The required data is measured using networks with a wide range of sensors. Nevertheless, in the context of smart cities no existing solution for traffic flow detection and vehicle classification is able to guarantee high classification accuracy, low deployment and maintenance costs, low power consumption and a weather-independent operation while respecting privacy. In this paper, we propose a radio-based approach for traffic flow detection and vehicle classification using signal attenuation measurements and machine learning algorithms. The results of comprehensive measurements in the field prove its high classification success rate of about 99%.

## I. Introduction

Intelligent traffic control mechanisms aim to reduce traffic jam occurrences and travel times as well as the $CO_2$ output. These objectives are also key components for future smart cities [1]. In order to achieve these goals, knowledge about the current traffic flow needs to be obtained at chosen measurement locations. Apart from the intelligent traffic control, there are further application fields which can benefit from traffic flow monitoring. For example, smart parking or toll monitoring systems can aggregate such data about the type of a vehicle for providing information about the parking space capacity or for calculating correct toll fees. In this paper, we propose a radio-based system which leverages the attenuation of radio signals for traffic flow detection and vehicle classification using machine learning algorithms. In contrast to other existing approaches, the proposed system is cost-efficient, easy to install and does not raise privacy-related issues because it is not based on the evaluation of camera images. Future smart cities could easily integrate it in delineator posts or traffic lights. The remainder of this paper is as follows. After discussing the related work (cf. Sec. II), we present the setup of our radio-based detection and classification system in Sec. III. In the next section, the system model of our multi-methodical solution approach and its individual components are explained (cf. Sec. IV). In Sec. V, detailed results for the classification accuracy are discussed, which compare the suitability of different machine learning algorithms and features for the defined problem statement. Moreover, the impact of ground-reflected radio waves regarding the Received Signal Strength Indicator (RSSI) is analyzed. Finally, the results show the high efficiency of the proposed approach and its suitability for being used in future smart cities.

## II. Related Work

Traffic flow detection has been a topic of scientific interest for a long time. In [2] a comparison of several approaches is presented. In recent years, the data of an increasing number of various detection systems is aggregated building a multi-functional data-driven ITS [3]. For this purpose, a wide range of different sensor and detection techniques with specific advantages and disadvantages is used. A widespread approach for vehicle detection and classification are camera-based systems, which achieve a high classification success rate. Normally, several cameras are needed in those systems to analyze the scenario from different angles and perspectives. In contrast, an enhanced visual system which is able to categorize vehicles into various vehicle classes using a single camera is presented in [4]. Regardless of the lower number of cameras, the presence of camera-based systems still raises a lot of additional effort in terms of installation, maintenance and also privacy-related problems in real-world scenarios. Furthermore, the success rate of these systems significantly decreases by challenging weather conditions. Among pure camera-based detection systems, there are approaches using laser scanners [5], acoustic sensors [6], magnetometers [7] or accelerometers [8]. While the former three approaches suffer from similar disadvantages like camera-based detection systems, the latter one requires possibly heavy construction works (pavement cut, etc). A more convenient approach using a single and portable magnetic sensor is presented in [9]. The sensor node is able to perform a stand-alone detection and classification of vehicles in real time with 99% detection and 60% classification success rate. In [10] the authors present an approach based on a single battery-powered magnetometer classifying vehicles into three categories. By using Support Vector Machine (SVM) the system reaches a classification accuracy of about 87%. Multiple and spatially separated magnetic sensors are used in the classification system proposed in [11]. Due to the redundancy of the sensors, the system is able to detect driving behavior in terms of right-turning or straight-driving vehicles. Apart from this, the system achieves a classification accuracy of only 83% when using SVM. In order to deal with the previously mentioned disadvantages and to achieve high classification success rates, we propose a detection and classification system which is integrated in present traffic infrastructure (e.g., delineator posts) leveraging the attenuation of radio signals caused by vehicles passing the setup. Essential preparatory work with regard to the system setup has been done in a previous project to detect wrong way drivers on motorways [12], which has a detection rate of over 99% for different traffic conditions. Therefore, we adopted the symmetrical setup with spatially separated transmitting and receiving units to extend the system to a high-precision detection and classification system for parking space accounting on motorway service areas based on the analysis of signal attenuation. The fact that vehicles can be considered as obstacles in the signal propagation path is elaborated in [13]. Subsequent to the evaluation of the signal attenuation, the vehicle classification is done with the help of different state-of-the-art machine learning algorithms. A comparison of Logistic Regression (LR), Neural Networks (NNs) and SVMs for magnetometer- and accelerometer-based vehicle classification is presented in [14]. The different algorithms achieve similar classification success rates of about 93%. To further enhance the success rate of the proposed classification system, we additionally use realistic ray tracing simulations to evaluate the most suitable system settings (e.g., antenna characteristics) and special situations. In general, ray tracing simulations are a popular method to generate close to reality data, especially in exceptional situations. The high suitability of ray tracing simulations in the context of vehicular communication is asserted in [15]. The authors give an accuracy comparison of channel-sounder measurements and ray tracing simulations assigning a high agreement between simulation and measurement data.

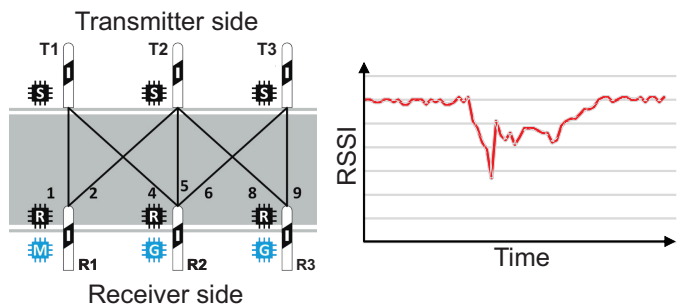

Fig. 1. Example application scenario: structure of the system setup consisting of transmitters and receivers and the effect of passing vehicles on the RSSI of radio links

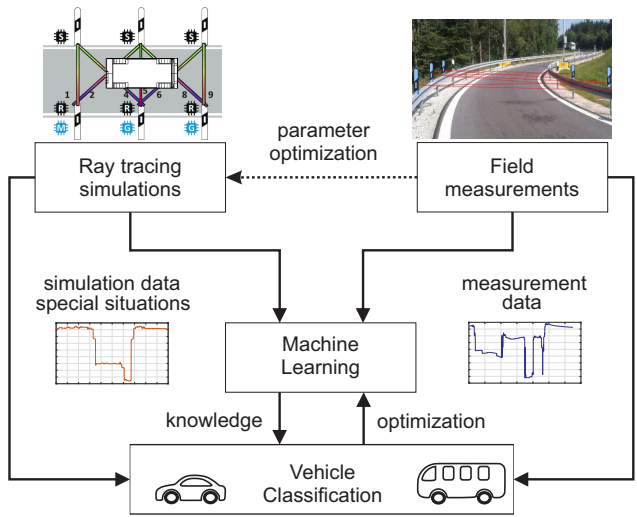

Fig. 2. System model of the proposed solution approach using ray tracing simulations, measurements in the field and machine learning algorithms.

## III. SETUP OF THE RADIO-BASED DETECTION AND CLASSIFICATION SYSTEM

The structure of the proposed detection and classification system is illustrated in Fig. 1. It primarily consists of two wireless key elements: transmitting and receiving units, which are positioned on opposite sides of a road spanning a radio field. It should be pointed out that each transmitter has an individual subset of radio links with different receiving nodes. For example, the radio links numbered 1 and 4 are associated to transmitter 1, whereas the links named 6 and 9 are related to transmitter 3. Typically, the Line of Sight (LOS) propagation path for radio transmissions has the largest share of the total transmission power compared to None Line of Sight (NLOS) paths. Hence, the passage of vehicles through the classification system results in chronologically fluctuating RSSI traces caused by shadowing. Comprehensive measurements in the field have shown that different types of vehicles have specific RSSI fingerprints. In particular, we use this fact to perform a machine-based vehicle classification for different types of vehicles.

## IV. MULTI-METHODICAL APPROACH

For the total system, three different methodological approaches are brought together: field measurements, ray tracing simulations and machine learning algorithms (cf. Fig. 2). Data generated in field measurements is used for training and testing purposes as well as for the continuous optimization of the ray tracing simulation parameters. This parameter adjustment is necessary to achieve a high degree of compliance of data generated by simulations with measurement data. In contrast to field measurements, ray tracing simulations are used primarily to evaluate the most suitable system settings (e.g., antenna characteristics) and secondarily to train and test the vehicle classification procedure regarding special situations. For example, this includes vehicle types or driving behavior we have not observed in field measurements yet. In this way, the classification system can be better prepared on special situations. In the next step, we use the outcome of the vehicle classification procedure to recursively optimize our machine-based classification system. Finally, the overall classification process is verified by comprehensive measurements in the field.

TABLE I
PARAMETERS OF THE RAY TRACING SIMULATION

| Parameter | Value |
|---|---|
| Number of vehicle types | 11 |
| Simulation runs per vehicle type | 50 |
| Interval steps | 0.01 s |
| Operating frequency | 2.4 GHz |
| Transmitting power | 2.5 dBm |
| Antenna type | omni-/directional |
| Gain of directional antennas | 7.1 dBi |
| Azimuth | 60 deg |
| Downtilt | 5 deg |
| Range of transmitters height | 0.4 m - 1.2 m |
| Range of receivers height | 0.4 m - 1.2 m |
| Step size for height | 0.2 m |
| Distance between transmitters and receivers | 7 m |

## A. Ray Tracing Simulations

In order to analyze the effects of different measurement system settings (e.g., characteristics of antennas) and to find the most suitable parametrization for the live system, we use the tools *WallMan* and *ProMan* of the ray tracing simulation tool suite *WinProp* and the Computer-aided Design (CAD) tool *Sketchup* for creating and adapting simulation scenarios with highly detailed vehicle models. On the basis of these simulation results, we have evaluated the most appropriate antenna settings (e.g., antenna type and height for installation). Fig. 3 illustrates an example simulation scenario based on the proposed system model consisting of multiple transmitters and receivers, which is passed by a SUV. Here, the colored box represents the location-dependent received power of the signal transmitted by transmitter 2. Apparently, the passage of vehicles has significant effects on the received signal strength.

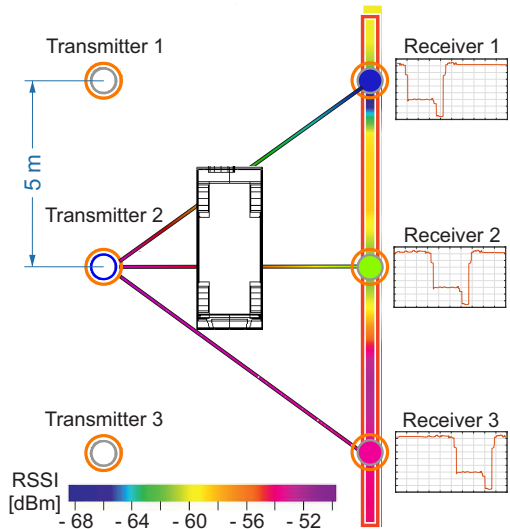

Fig. 3. Example ray tracing scenario illustrating the shadowing effects caused by a SUV passing the proposed classification system.

TABLE II
PARAMETERS OF THE FIELD MEASUREMENT SETUP

| Parameter | Value |
|---|---|
| Covered area | 10 m x 5 m |
| Structure of senders and receivers | symmetric |
| Number of senders | 3 |
| Number of receivers | 3 |
| Types of signal paths | direct, diagonal |
| Number of signals per receiver | 2-3 |
| Operating frequency | 2.4 GHz |
| RF power | 2.5 dBm |

Tab. I lists relevant parameters used for the simulations. After modeling the simulation scenario, each of its object is provided with appropriate material and motion properties, respectively. Both parameters are of high relevance regarding a realistic simulation of the signal characteristics at the receiving unit. With reference to the movement behavior, it is possible to distinguish between speed, direction of motion and relative distance to transmitter and receiver, respectively. Depending on the simulation scenario, a few motion aspects are limited regarding their range of values. For example, it is unlikely that vehicles drive with very high speeds and in the wrong direction through a parking space. Nevertheless, some of these scenarios are also simulated to evaluate the most suitable settings for high classification success rates of the live system in the case of special situations.

## B. Machine Learning

For the vehicle classification we use different state-of-the-art machine learning algorithms: k-Nearest Neighbor (k-NN) and SVM. These algorithms are trained and tested with the help of raw data containing chronological RSSI traces from field measurements and simulations. With this approach, the knowledge for classification is consecutively enhanced with a large set of real-world data as well as a small portion of simulation data. Beside the RSSI traces length information is used as second feature for classification. The training of the classification system has been done supervised with the two labels *passenger car* and *truck*. To control the performance of the classification process a standard five-fold cross-validation was used.

## C. Measurements in the Field

For validating the classification success rate of the proposed machine learning-based system, we use new data sets generated by field measurements. In order to achieve a high variety of vehicle types as well as a low influence of the environment on the classification, we have chosen a driveway for a parking space next to the motorway A9 in Germany as location for our field measurement setup. The most important parameters of this setup are summarized in Tab. II. Basically, the setup consists of a total of six delineator posts equipped with directional antennas, microcontroller boards for signal processing and a power supply. While one half of those delineators serves as transmitters, the remaining delineators are used as receivers. In addition to the signal reception, the receivers are also responsible for signal processing and forwarding of the processed data to the master gateway. After collecting all data, the master gateway performs the main part of the whole classification procedure.

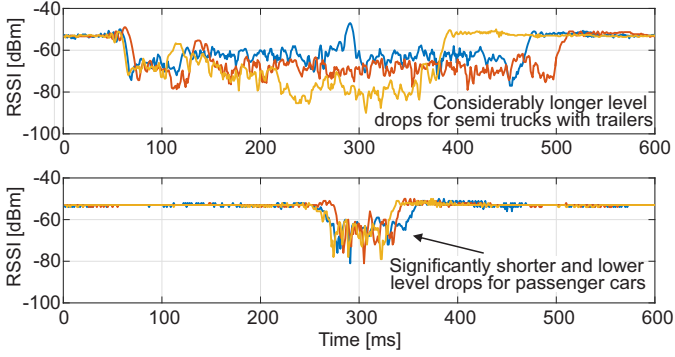

Fig. 4. Example field measurement data of three trucks and three passenger cars.

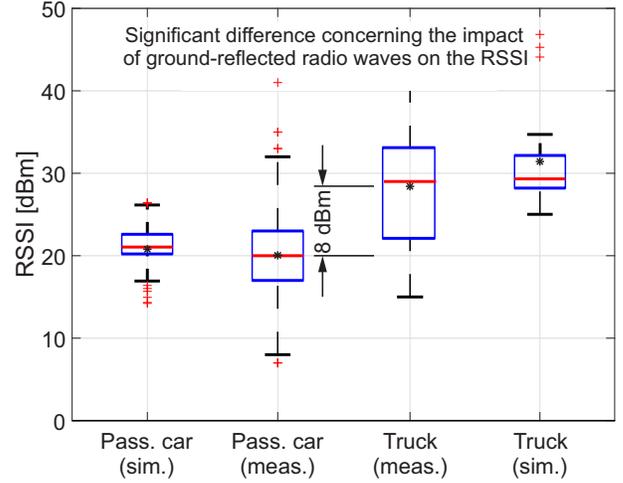

Fig. 5. Analysis of the influence of ground-reflected radio waves on the RSSI for passenger cars and trucks based on measurement and simulation data.

## V. RESULTS OF THE PERFORMANCE EVALUATION

In this section, we present the results achieved with the proposed classification system. First, we examine the significance of ground-reflections on the received signal strength with the help of measurement and simulation data. Then, we present the results using the features length information and RSSI traces for classification. Finally, we consider the classification success rate of the proposed system as our main key performance indicator.

### A. Impact of Ground-reflected Radio Waves

Subsequently, the results of the proposed classification system for use of field measurement data are presented. Fig. 4 shows example RSSI traces caused by the passage of trucks and passenger cars. Obviously, the developing of the RSSI traces differs in terms of duration and magnitude of signal dropout depending on the vehicle type passing the measurement system. In particular, trucks cause a significantly higher and considerably longer drop of the RSSI levels compared to passenger cars. In addition, the temporal developing indicates whether a truck with or without a trailer is passing the classification system. For example, one trace in the upper part of Fig. 4 illustrates the passage of a semi truck with a trailer. The reason for the temporary peak of signal strength at the time of about 300 ms is the free space between the semi truck and the trailer. At this point, there is almost a LOS signal path between transmitter and receiver resulting in a significantly higher RSSI level. To check the suitability of RSSI traces as a feature for the classification process, we have analyzed the impact of ground-reflected radio waves across the street on the received signal strength. Fig. 5 shows the magnitudes of RSSI level drops due to the shadowing caused by passenger cars and trucks for measurement and simulation data, respectively. Apparently, the mean drop of signal strengths caused by passenger cars is about 8 dBm higher compared to the one caused by trucks (cf. measurement data in Fig. 5). This results from the significantly different shapes and distances of the car bodies of various vehicle types and the surface of a street. As a consequence, the RSSI traces of radio links across the street differ for various vehicle types due to varying impacts of ground-reflections caused by the street. This assumption is confirmed with the help of simulation results also shown in Fig. 5. Finally, these results reveal the high suitability of RSSI traces as feature for vehicle classification.

### B. Classification Results using Field Measurement Data

In order to achieve a high classification accuracy for various types of vehicles, the machine learning algorithms have been trained and tested via five-fold cross validation. By using RSSI traces as exclusive feature, the cross validation leads to an accuracy of $98.68\% \pm 0\%$ for k-NN and $98.68\% \pm 0.31\%$ for SVM (cf. Tab. IV). If length information is used as an additional feature, an accuracy of $99.56\% \pm 0\%$ for k-NN and $99.47\% \pm 0.20\%$ for SVM is achieved. Tab. III contains the results of the proposed classification system for the two labels passenger car and truck when using only aggregated length information, the RSSI traces as exclusive feature or the combination of both features. The classification results have been evaluated with *MATLAB* and validated with the help of *RapidMiner* [16].

TABLE III
CLASSIFICATION SUCCESS RATE OF K-NN AND SVM REGARDING THE LABELS PASSENGER CAR AND TRUCK WITH FIELD MEASUREMENT DATA

| | | Length | | RSSI Traces | | | Length & RSSI Traces | | |
|---|---|---|---|---|---|---|---|---|---|
| Label | Vehicle type | Test samples | Rec. rate | Test samples | k-NN | SVM | Test samples | k-NN | SVM |
| Passenger car | Passenger car | 541 | 100.0% | 60 | 100.0% | 100.0% | 60 | 100.0 % | 100.0% |
| | Small van | 12 | 100.0% | 12 | 100.0% | 91.67% | 12 | 100.0% | 100.0% |
| | Van | 100 | 92.00% | 20 | 100.0% | 100.0% | 20 | 100.0% | 100.0% |
| | Transporter | 107 | 85.98% | 27 | 96.30% | 96.30% | 27 | 96.30% | 92.59% |
| Truck | Bus | 6 | 100.0% | 6 | 100.0% | 100.0% | 6 | 100.0% | 100.0% |
| | Truck | 503 | 94.04% | 103 | 98.06% | 99.03% | 103 | 100.0% | 100.0% |
| Overall success rate | | 1269 | 95.82% | 228 | 98.68% | 98.68% | 228 | 99.56% | 99.12% |

TABLE IV
CROSS VALIDATION: CLASSIFICATION SUCCESS RATES OF k-NN AND SVM REGARDING THE RECOGNITION OF THE LABELS PASSENGER CAR AND TRUCK USING RSSI TRACES AS EXCLUSIVE FEATURE.

| Training set | Test set | k-NN | SVM |
| --- | --- | --- | --- |
| S2, S3, S4, S5 | S1 | 98.68% | 98.68% |
| S1, S3, S4, S5 | S2 | 98.68% | 98.68% |
| S1, S2, S4, S5 | S3 | 98.68% | 98.25% |
| S1, S2, S3, S5 | S4 | 98.68% | 98.68% |
| S1, S2, S3, S4 | S5 | 98.68% | 99.12% |

Due to the symmetrical structure of the proposed system, a highly accurate determination of vehicle length information is possible, which can also be used for vehicle classification. However, for a few vehicle types (e.g., transporters) the sole evaluation of length information results in comparably low classification success rates. The reason for this is the wide variety of vehicles associated with only two labels. For example, there are transporters differentiating in terms of shape and size. Therefore, the members of a vehicle type may differ significantly from each other. In comparison to the sole evaluation of length information (95.82% total accuracy), overall classification success rates of 98.68% can be achieved by using RSSI traces. In order to further improve the classification success rate, the combination of both features is used for classification (cf. Tab. III). Apparently, the overall classification success rates increase from 98.68% to 99.56% (k-NN) and from 98.68% to 99.12% (SVM). In particular, the advantage of this approach is revealed with regard to the recognition of trucks. If RSSI traces are used exclusively, the classification system achieves an accuracy of 98.06% (k-NN) and 99.03% (SVM). Instead, the usage of both features leads to classification success rates of 100% when using k-NN and SVM, respectively. Finally, the overall classification success rates are slightly improved by combining both features.

## VI. CONCLUSION

In this paper, we presented a radio-based approach for traffic flow detection and vehicle classification which combines ray tracing simulations, machine learning and measurements in the field. First, the impact of ground-reflected radio waves on the RSSI traces were examined revealing their high suitability as feature for our classification system. The results proved the high performance of the proposed classification system with regard to the classification of two labels. By combining the two features RSSI traces and vehicle length information the classification system achieves an accuracy of more than 99%. Moreover, the proposed classification system has low deployment and maintenance costs recommending it for obtaining relevant data in a smart city context. In future work, we will increase the accuracy of the proposed system with the help of sensor fusion and the evaluation of further signal characteristics even more.

ACKNOWLEDGMENT

Part of the work on this paper has been supported by the German Federal Ministry for Economic Affairs and Energy as part of the cooperation project between Wilhelm Schröder GmbH, TU Dortmund and FH Dortmund (grant agreement number ZF4038101DB5), and by Deutsche Forschungsgemeinschaft (DFG) within the Collaborative Research Center SFB 876 "Providing Information by Resource-Constrained Analysis", project B4 "Analysis and Communication for Dynamic Traffic Prognosis".